\documentclass[9pt,twocolumn,twoside]{pnas-new}
\setboolean{displaywatermark}{false}

\usepackage{natbib}
\usepackage{xr}
\usepackage{bm}
\usepackage{hyperref}
\usepackage{xcolor}

\templatetype{pnasresearcharticle} % Choose template 
\newcommand{\Wo}{\mathit{Wo}}
\newcommand{\A}{\mathit{A}}

\title{Nonlinear hydrodynamic instability and turbulence in pulsatile flow}

\author[a,b,c,1,2]{Duo Xu}
\author[a,1]{Atul Varshney} 
\author[a]{Xingyu Ma}
\author[d]{Baofang Song}
\author[a]{Michael Riedl} 
\author[b]{Marc Avila}
\author[a,2]{Bj{\"o}rn Hof}

\affil[a]{Institute of Science and Technology Austria, Am Campus 1, Klosterneuburg 3400, Austria}
\affil[b]{University of Bremen, Center of Applied Space Technology and Microgravity  (ZARM), 28359 Bremen, Germany}
\affil[c]{Friedrich-Alexander-Universit\"{a}t Erlangen-N\"{u}rnberg, Erlangen 91058, Germany}
\affil[d]{Tianjin University, Center for Applied Mathematics, Tianjin 300072, China}

\leadauthor{Xu} 

\significancestatement{
The inner lining of blood vessels, the endothelium, is shear sensitive. Fluctuating shear stresses and disordered flow are responsible for cellular dysfunction, leading to the development of atherosclerotic lesions. We here identify a nonlinear hydrodynamic instability that gives rise to disordered motion in the parameter range of cardiovascular flow. During flow deceleration small geometrical imperfections trigger a helical vortex pattern that subsequently breaks down into bursts of turbulence. The resulting fluctuating shear stress level drops abruptly during the acceleration where flow relaminarization sets in. The observed instability occurs at much lower flowrates than either the linear instability or the classical route to turbulence. Our study shows that disordered motion is more common in pulsatile/cardiovascular flows than previous stability considerations suggest.
}

\authorcontributions{M.A. and B.H. designed the research; D.X., A.V., X.M. and M.R. performed research; D.X., A.V. and M.R. analyzed data; and D.X., A.V., M.A. and B.H. wrote the paper. }
\authordeclaration{The authors declare no conflict of interest.}
\equalauthors{\textsuperscript{1}D.X. and A.V. contributed equally to this work.}
\correspondingauthor{\textsuperscript{2}To whom correspondence should be addressed. E-mail: bhof@ist.ac.at or duo.xu@zarm.uni-bremen.de.}

\keywords{hydrodynamic instability $|$ transition to turbulence $|$ pulsatile flow $|$ (non-)Newtonian fluids} 

\begin{abstract}
Pulsating flows through tubular geometries are laminar provided that velocities are moderate. This in particular is also believed to apply to cardiovascular flows where inertial forces are typically too low to sustain turbulence. On the other hand flow instabilities and fluctuating shear stresses are held responsible for a variety of cardiovascular diseases. Here we report a nonlinear instability mechanism for pulsating pipe flow that gives rise to bursts of turbulence at low flow rates. Geometrical distortions of small, yet finite amplitude are found to excite a state consisting of helical vortices during flow deceleration. The resulting flow pattern grows rapidly in magnitude, breaks down into turbulence, and eventually returns to laminar when the flow accelerates. This scenario causes shear stress fluctuations and flow reversal during each pulsation cycle. Such unsteady conditions can adversely affect blood vessels and have been shown to promote inflammation and dysfunction of the shear stress sensitive endothelial cell layer.
\end{abstract}

\dates{This manuscript was compiled on \today}
\doi{\url{www.pnas.org/cgi/doi/10.1073/pnas.XXXXXXXXXX}}

\begin{document}

\maketitle
\thispagestyle{firststyle}
\ifthenelse{\boolean{shortarticle}}{\ifthenelse{\boolean{singlecolumn}}{\abscontentformatted}{\abscontent}}{}

\dropcap{B}lood vessels react to hemodynamic forces and in particular the vessels' inner layer, the endothelium, is highly shear sensitive. Fluctuating flow and low wall shear stress levels cause inflammation of the endothelium, which in turn can lead to the development of atherosclerosis lesions \cite{nerem_role_1980, cunningham_role_2005, gimbrone_endothelial_2016}. However, the hydrodynamic instabilities responsible for fluctuations and varying shear stress levels are often unknown. 
Already for the simpler case of steadily driven flow through a straight pipe it is non-trivial to predict if the fluid motion will be smooth and laminar or highly fluctuating and turbulent. In that case the laminar state is linearly stable yet turbulence arises as a result of finite amplitude perturbations provided that the Reynolds number ($Re$) is sufficiently large. It is characteristic for the `subcritical instability' scenario that turbulence does not appear globally but only at the location where the laminar flow is perturbed and here a localized patch, a `puff', of turbulence is formed \cite{Wygnanski73,Hof06, Hof08, Avila10}. Puffs have a constant size and travel downstream at approximately the bulk flow velocity. In steady pipe flow turbulence never spreads upstream and this instability is hence of convective nature \citep{Huerre90, Chomaz05}.

Pulsatile flows are more complex and governed by two additional control parameters, i.e. the pulsation amplitude and frequency (Womersley number). Depending on parameters the primary instability encountered differs qualitatively. For predominantly oscillatory flows, i.e. flows with small or no mean flow component, the flow becomes linearly unstable even though the cycle averaged Reynolds number vanishes in this limit. This linear instability has been extensively investigated and is well understood \cite{merkli_transition_1975, davis_stability_1976, hall_linear_1978, kerczek_instability_1982, blennerhassett_linear_2002}. In contrast the flow in blood vessels is pulsatile, i.e. it is dominated by the mean flow and the oscillatory component is smaller. Although the aforementioned linear instability is also found in this case, it only occurs at very high flow speeds \citep{Thomas11}. The corresponding critical Reynolds number lies far above the values encountered in blood flows and hence this transition threshold is not relevant for cardiovascular flow.
In addition, the aforementioned subcritical instability to turbulent puffs persists to pulsatile flow \citep{Xu17,Xu18}. In cardiovascular flows, it is typically assumed that turbulence sets in at similar $Re$ as in steady pipe flow \cite{davis_stability_1976}. Hence for $Re<2000$ flows are deemed laminar while above transition may occur \cite{Avila11}. 

In large arteries, Reynolds numbers can reach peak values considerably larger than this limit. However mean values even in the aorta typically do not exceed $Re=2000$. In addition, more recently it has been shown that for high pulsation amplitudes and low frequencies the transition to puffs is delayed \cite{Xu17, Xu18}. In cardiovascular flows, on the other hand instabilities are commonly observed during flow deceleration \cite{ku_blood_1997, chien_effects_2008, davies_hemodynamic_2009}. Furthermore, it is unclear how geometrical deviations from the generic straight pipe case (like bends, unevenness and junctions) affect the stability of the flow.

In the following we report a subcritical instability specific to pulsating flow. This nonlinear instability sets in during flow deceleration, downstream of small imperfections of the pipe, such as bends or protrusions. Initially, a helical wave arises which subsequently breaks down into turbulence, and fluctuation levels rise before they abruptly drop during the accelerating phase where flow relaminarization sets in. This helical instability is observed at $Re$ as low as $1000$, a threshold that is surpassed in a variety of larger vessels. We show that the observed mechanism is generic for pulsatile flow and the helical wave corresponds to the most amplified perturbation of the linearized equations. 

\begin{figure*}[htbp]
	\centering
	\includegraphics[width = 1\textwidth]{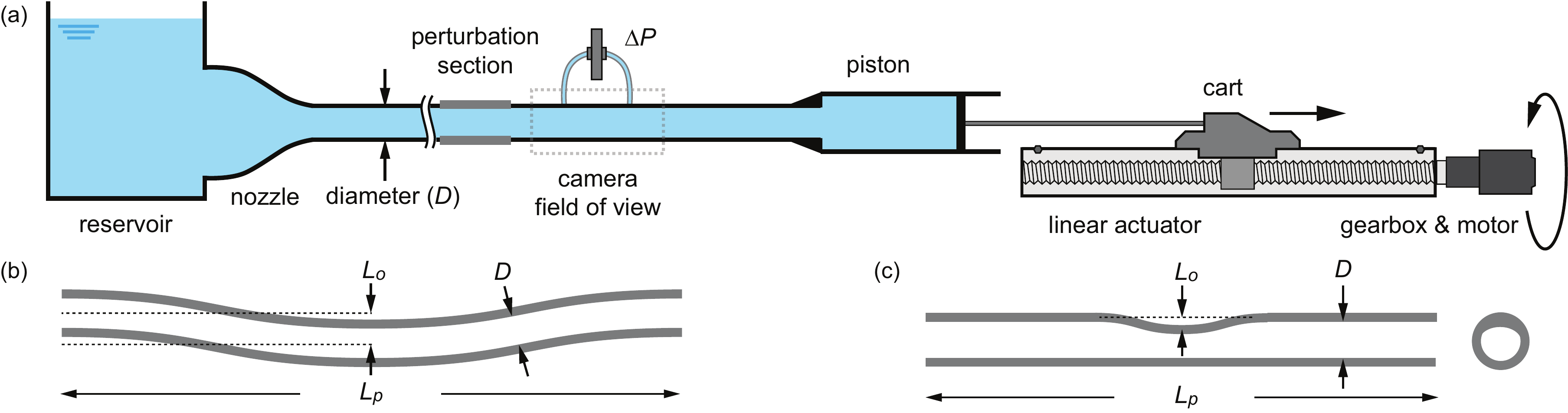}
	\caption{\label{fig:facility} {{(Color online) (a) Sketch of the pulsatile pipe flow setup. The dashed rectangle marks the measurement location where pressure and visualization measurements were carried out. The flow is from left to right. The perturbation methods are sketched in (b) and (c) where $D$ is the pipe inner diameter, $L_p$ the length of perturbation section and $L_o$ the offset: (b) curved segment perturbation where $L_p=7D$ and the offset ranges from $0D$ (natural transition) to $0.45D$; (c) constriction perturbation (mimicking unevenness) where $L_p=5.6D$ and the offset ranges from $0.14D$ to $0.7D$. The right sub-panel shows the cross-sectional view.}}}
\end{figure*}
\begin{figure}[htb]
	\centering
	\includegraphics[width=0.9\linewidth]{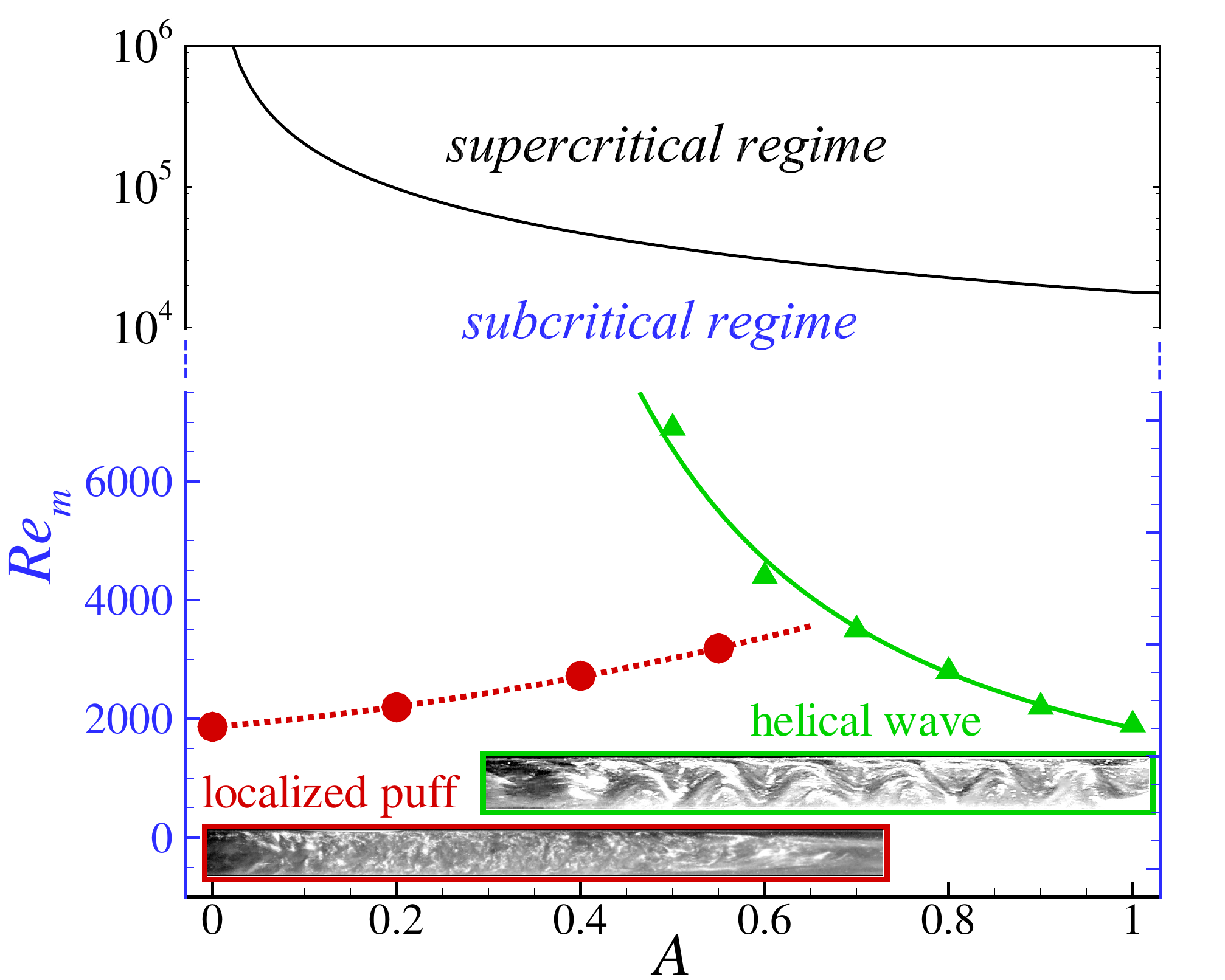}
	\caption{{(Color online) The threshold for the onset of puffs is given by the  red dotted line. That for the onset of the helical wave instability is given by the green solid line. The Womersley number is held fixed at $\Wo=5.6$. The upper part of the figure (note the scale is altered to be logarithmic) shows the linear instability threshold (black curve) which only sets in at $Re_m$ much larger than those discussed in this study. The inset shows flow visualization images at $t/T\approx0.68$: the top panel shows the helical wave pattern and the bottom panel shows a puff. The flow in both cases is from left to right. }}	
	\label{fig:puff_wavy}
\end{figure}

\section*{Results}

\begin{figure*}[htb]
	\centering
	\includegraphics[width=1\linewidth]{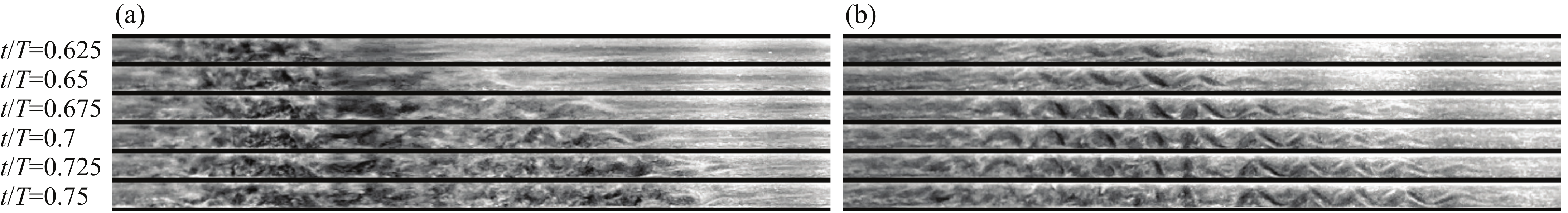}
	\caption{{Visualization of transition to turbulence} in pulsatile pipe flow in a space-time diagram at $(Re_m, \Wo, \A)=(2200, 5.6, 0.85)$: (a) The evolution of a puff which grows in the streamwise direction while its upstream interface is approximately stationary. (b) Evolution of the helical instability. The helical wave spreads downstream as well as upstream. The flow in (a) and (b) is from left to right.}	
	\label{fig:visualization}
\end{figure*}

\begin{figure}[!h]
	\centering
	\includegraphics[width=1\linewidth]{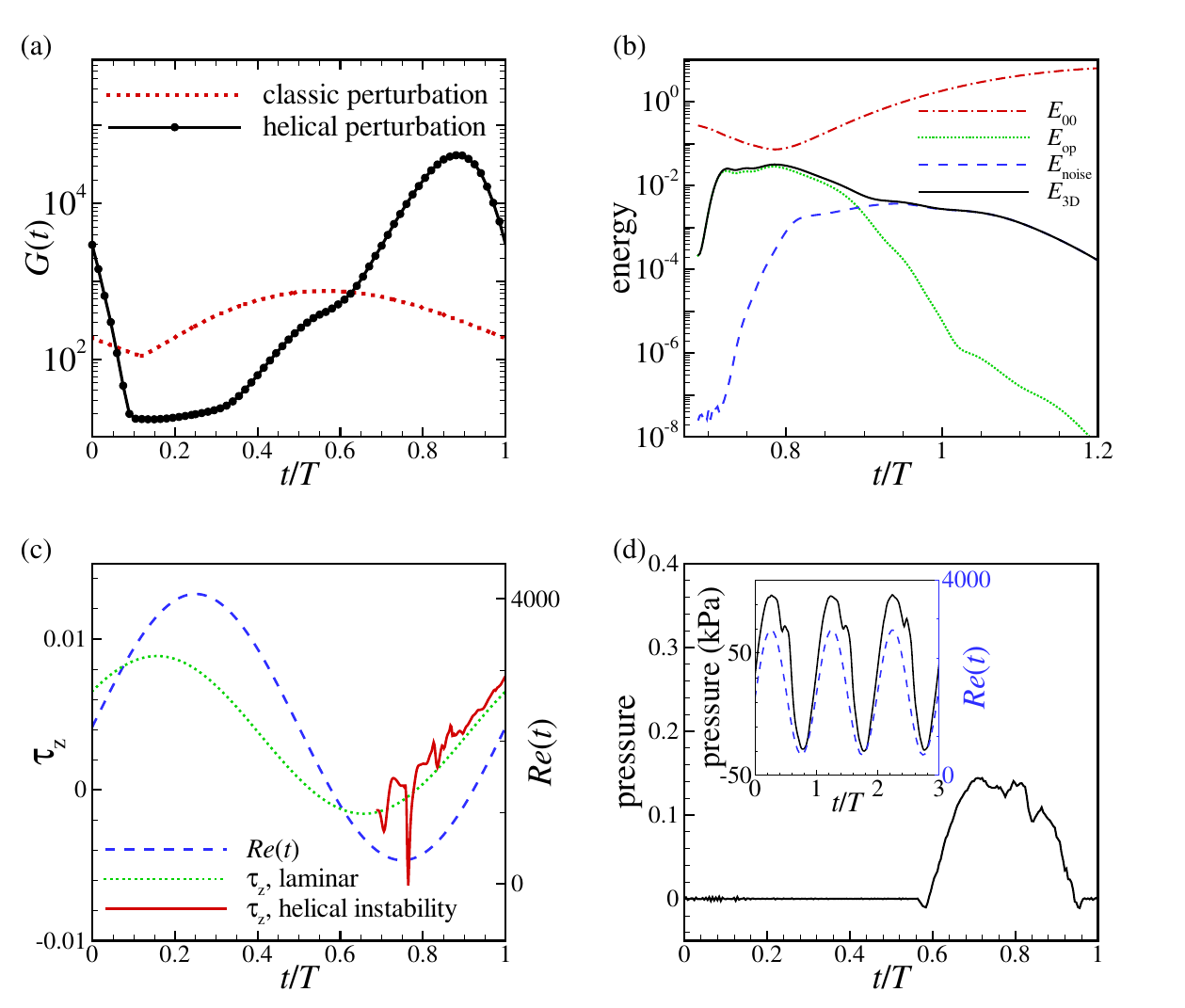}
	\caption{{(Color online) (a) Optimal linear energy growth $G(t)$ of disturbances at $(Re_m,\Wo,\A)=(2200,5.6,0.85)$ for the classic perturbation (streamwise independent, dotted line) and helical perturbation (solid line).  (b) Direct numerical simulation of transition in a pipe of $12D$ length disturbed with the optimal helical perturbation (for $1.5D$ wavelength)  and superposed three-dimensional noise. Shown are time series of the kinetic energy of the spatially averaged flow profile ($E_\text{00}$) and the three-dimensional component of the disturbance ($E_\text{3D}$), i.e.\ of those Fourier modes with $k\ne0$ and $m\ne0$. The latter is further decomposed into the part corresponding to the  optimal helical perturbation ($E_\text{op}$) and the rest ($E_\text{noise}$). 
	(c) Time series of fluid wall shear stress $\tau_z$ exerting on pipe wall at a fixed location, together with the instantaneous Reynolds number $Re(t)$, in the direct numerical simulation at $(Re_m,\Wo,\A)=(2200,5.6,0.85)$. (d) The relative deviation in pressure from the corresponding laminar case for blood flow at $(Re_m,\Wo,\A)=(1140,4.0,0.5)$. The inset shows the time series of streamwise differential pressure (black solid line) together with the instantaneous Reynolds number (blue dash line) at $(Re_m,\Wo,\A)=(1700,5.9,0.58)$. The instability causes the smaller secondary peak in the pressure signal during flow deceleration.}}	
	\label{fig:time_friction_energy}
\end{figure}

\subsection*{Puff Turbulence (Experiment)}
Initial experiments were carried out in a rigid straight pipe with an inner diameter of $7~\mathrm{mm}$ and a total length of $12~\mathrm{m}$. The fluid was pulled through the pipe by a piston (see Fig.~\ref{fig:facility}). The piston speed was sinusoidally modulated imposing a cross--sectionally averaged flow velocity $U(t)=U_m + U_o\cdot \mathrm{sin}(2\pi f t)$, where $U_m$ is the mean flow speed, $U_o$ the oscillation component of the flow speed, $f$ the frequency and $t$ is the time. In accord with linear stability theory, the unperturbed flow remains laminar over the entire parameter regime investigated. In order to identify the flows susceptibility to finite amplitude perturbations, an impulsive jet of fluid could be injected through a small hole in the pipe wall, located $150D$ downstream of the pipe inlet. 
To visualize the flow structure, the water was seeded with reflective particles ({fishsilver}) and a light sheet was used to illuminate the mid cross section (radial-streamwise) of the pipe.
At sufficiently large $Re$ the perturbed flow develops into a turbulent puff which is then advected downstream. An example of a puff with its characteristic intense upstream interface and a gradual downstream interface is shown in Fig.~\ref{fig:puff_wavy}. 
Just like in steady pipe flow, puffs also have finite lifetimes in pulsatile flow. In order to determine the effect of flow pulsation on the puff transition threshold, we measured the puff survival rate for varying pulsation amplitude. While the frequency was held constant throughout (i.e. Womersley number, $\Wo=5.6$) for each selected pulsation amplitude, the Reynolds number was increased until puffs were first detected. Womersley number, pulsation amplitude and Reynolds number are defined as follows: $\Wo=0.5D\sqrt{2\pi f/\nu}$, $\A=U_o/U_m$ and $Re_m=U_mD/\nu$, where $D$ is the pipe diameter, $\nu$ is kinematic viscosity of the fluid. Whereas for low $Re$ all puffs decayed before the end of the pipe, at sufficiently large $Re$ all puffs would survive, and as a measure of the transition threshold we determined the Reynolds number where $50\%$ of puffs survive. For each pair of parameters ($\A$ and $Re_m$) lifetime statistics were based on a sample of $150$ puffs (see {ref.~\citenum{Xu17}} for further details about the general methodology). In Fig.~\ref{fig:puff_wavy} we plot the dependence of this chosen puff survival threshold on the pulsation amplitude. With increasing amplitude the puff transition (red curve) is delayed in accordance with ref.~\citenum{Xu17}. 

\subsection*{Helical Instability (Experiment)}
When the pulsation amplitude surpasses $0.7$, the above trend stops and the transition threshold begins to move to lower $Re_m$. Inspection of the flow structure shows that here instead of puffs a regular, helical vortex pattern is observed (see Fig.~\ref{fig:puff_wavy}). 
Unlike puffs this structure does not result from the injection of a jet at the perturbation location, but instead it was found to develop at a fixed pipe location at each cycle during flow deceleration (i.e. for $0.6\lesssim t/T\lesssim 0.75$ with period $T$) and it decays during acceleration (see Movie S$1$ in \emph{Supporting Information, SI}). Upon a further increase in the pulsation amplitude the instability threshold moves to smaller $Re_m$. The instability branch can also be continued to lower amplitudes ($A<0.7$), in this case we did not trigger puffs, but instead the Reynolds number was increased up to the point where the helical instability appeared naturally. 

Inspection of the pipe revealed that the pipe segment directly upstream of the location where the helical (wave) instability occurred, was slightly bent (with an axial misalignment of approximately $1~\mathrm{mm}$). When realigning the pipe, the helical instability could be postponed to larger $Re_m$, while further misalignment moved the instability threshold to lower $Re_m$.
To illustrate the structural and dynamic differences between puffs and the helical instability, we compare both at the same parameter values $(Re_m,\Wo,\A)=(2200, 5.6, 0.85)$. In one case the pipe segment was carefully aligned and a puff was triggered using the upstream injection perturbation, in the other case no puff was injected and the flow was perturbed by the upstream bend pipe segment. Both instances are shown for the flow deceleration phase in Fig.~\ref{fig:visualization}: The puff begins to spread in the downstream direction, while its upstream interface remains at the same location; Over the same part of the cycle, the helical instability gradually increases in amplitude and spreads down- as well as up- stream. The upstream propagation indicates that the instability is of absolute nature during part of the cycle, while the puff instability for the same parameters remains convective \citep{Huerre90, Chomaz05}.  

\begin{figure}[htbp]

 	\centering
 	\includegraphics[width=1\linewidth]{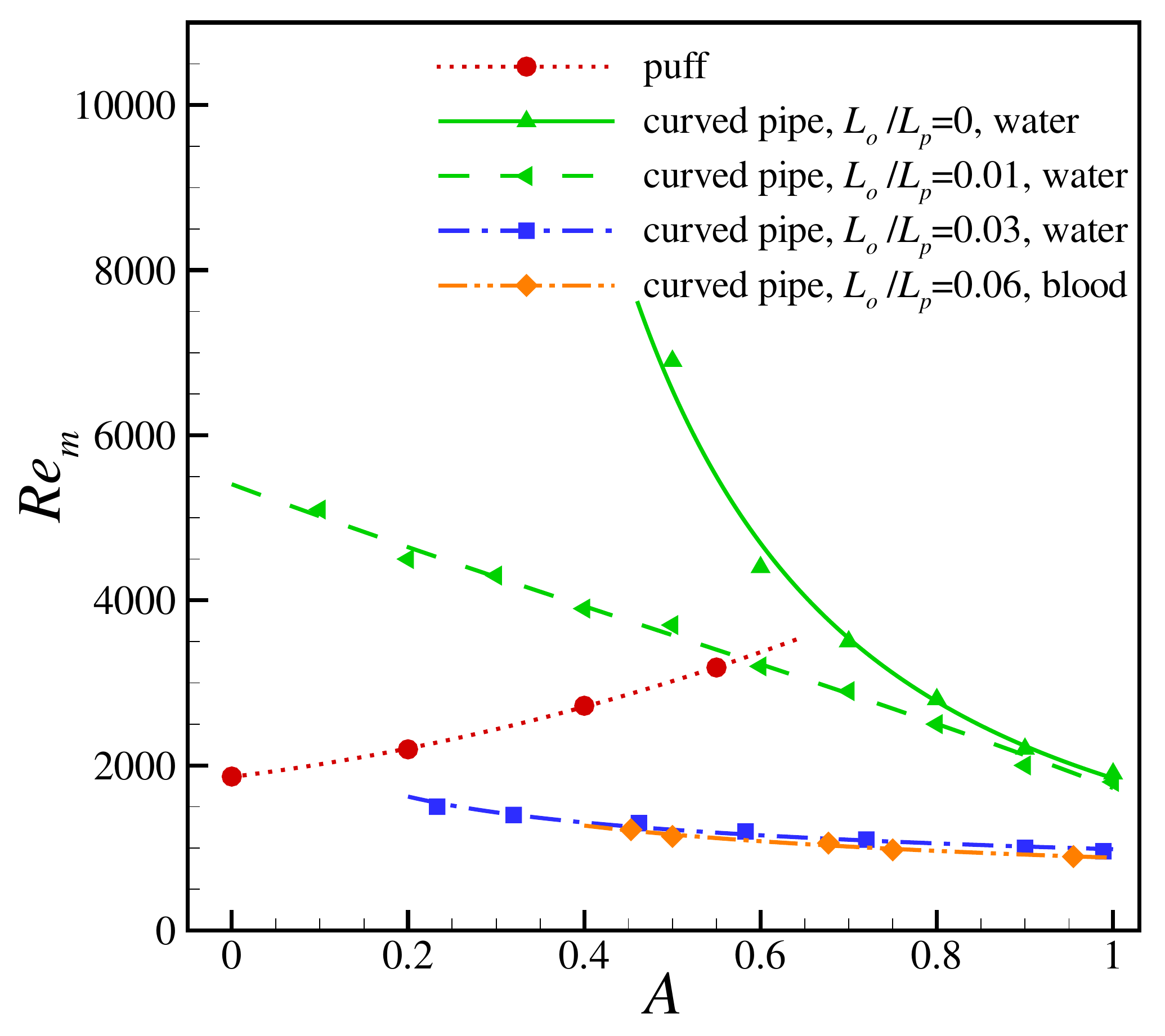}
 	\caption{{(Color online)  Onset of instability as a function of the pulsation amplitude for water (Newtonian) and blood (non-Newtonian). The pulsation frequencies (i.e. Womersley numbers) for the different data sets are as follows: red circles $\Wo=5.6$, green triangles $\Wo=5.6$ and blue squares $\Wo=5.9$. For the blood flow measurement (orange diamonds) $\Wo=4.0$.}}	
 	\label{fig:Re_A}
 \end{figure}

It should be noted that the misalignment considered above is only a fraction of a pipe diameter, and in the cardiovascular context virtually all blood vessels show deviations from the idealized straight pipe case, which are of that order or larger. To trigger the helical instability in a more controlled manner, we inserted a short pipe segment with a chosen moderate curvature (as sketched in Fig.~\ref{fig:facility}b, see \hyperref[sec:exp_method]{\textbf{Materials and Methods}} for details), while keeping the rest of the pipe straight and well aligned. With a more strongly curved pipe segment, the instability occurs at considerably lower $Re_m$ (see green curves in Fig.~\ref{fig:Re_A}) and again the transition threshold decreases with $\A$. 
These findings suggest that the helical instability, just like the instability to turbulence in steady flow, results from a perturbation of finite amplitude. While the transition in steady pipe flow is characterized by a double threshold \citep{Grossmann00}, i.e. both the amplitude of the perturbation and the Reynolds number have to be large enough, the helical instability has a triple threshold. Here in addition to the perturbation amplitude and the Reynolds number also the pulsation amplitude has to be sufficiently large. Moreover, the types of disturbance that trigger the helical instability differ from those triggering puffs.  

\begin{figure}[htb]
	\centering
	\includegraphics[width=1\linewidth]{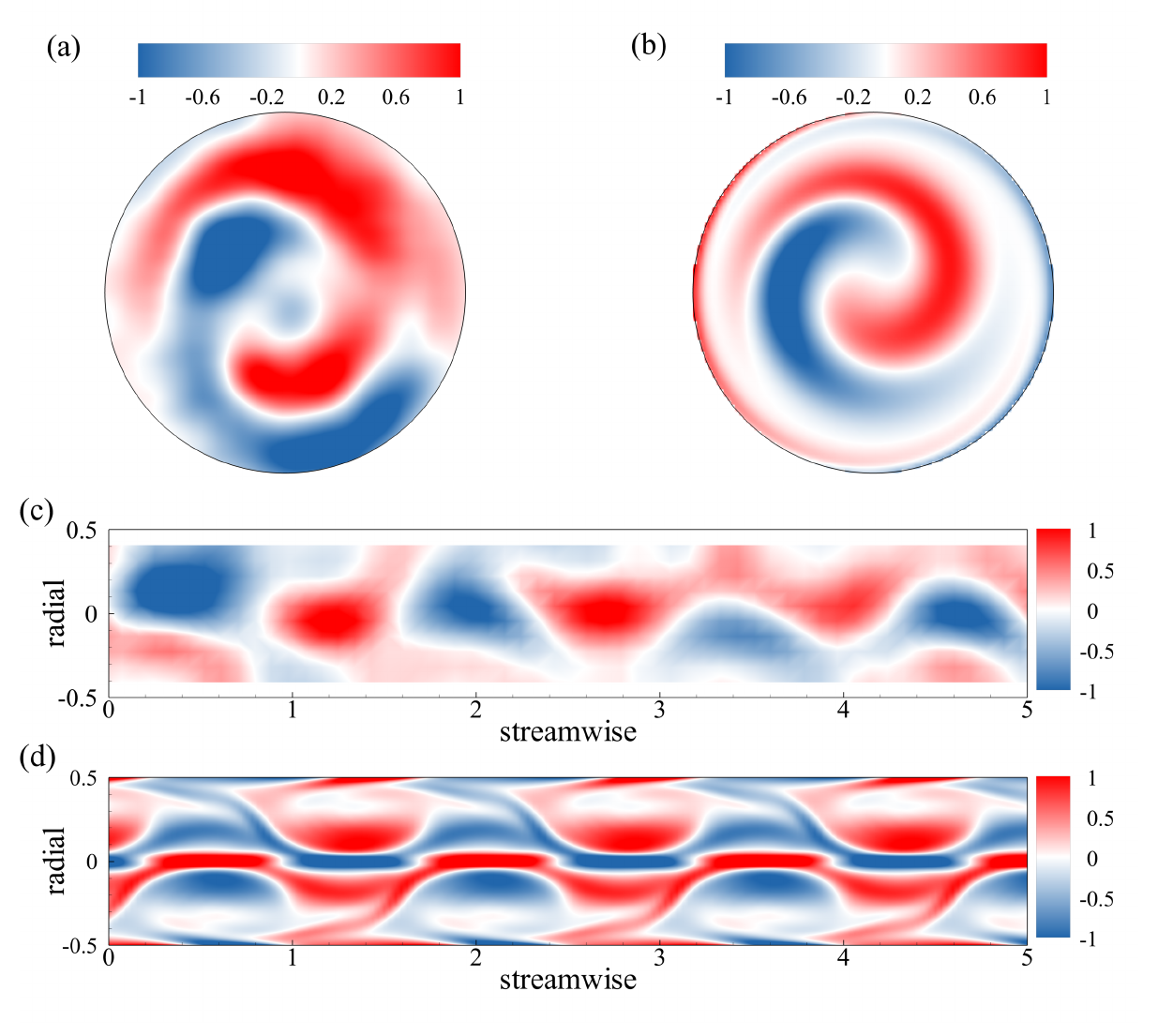}
	\caption{(Color online) Colormap of the streamwise vorticity in a radial-azimuthal cross-section of the pipe from experiments (a) and numerical simulations (b). The colormap of the spanwise vorticity in a radial-streamwise plane from experiments (c) and numerical simulations (d). In both cases, a pipe segment of $5D$ is shown. The experiment and the direct numerical simulation were both carried out at $(Re_m,\Wo,\A)=(2200,5.6,0.85)$, and the snapshots were taken at $t/T\approx0.7$. }	
	\label{fig:PIV_DNS_contour}
\end{figure}

\subsection*{Helical Instability in Simulation}
To elucidate the origin of the instability, we carried out numerical simulations of the Navier--Stokes equations. Albeit the laminar flow is linearly stable over the parameter range studied in the experiments, this does not preclude the possibility that perturbations can grow over part of the pulsation cycle, as long as they experience a net decay over the full cycle \cite{kerczek_instability_1982,Tsigklifis17}. We determined the optimal perturbations of pulsating pipe flow by performing a linear non-modal transient growth analysis with an adjoint-based method (see \hyperref[sec:num_method]{\textbf{Materials and Methods} for technical details}). As shown in Fig.~\ref{fig:time_friction_energy}a, the energy of infinitesimal perturbations can be amplified by more than four orders in magnitude during part of the cycle. Interestingly, the optimal perturbation has a helical shape and yields its maximum energy amplification toward the end of the deceleration phase. Overall, this helical perturbation dominates during the deceleration phase, and it has an optimal azimuthal wavenumber $m=1$ and an optimal wavelength of about $3D$, whereas the classic optimal perturbation of steady pipe flow  \cite{SchmidHenningson_springer2001} has also $m=1$, but is streamwise independent. The latter is also relevant to pulsatile pipe flow and dominates in the acceleration phase, but featuring much lower amplification factors than the helical perturbation in the deceleration phase. Note that beneath the solid line in Fig.~\ref{fig:time_friction_energy}a there are several families of highly amplified (suboptimal) helical perturbations parameterized by the axial wavelength. 

In order to compare to experiments, we carried out direct numerical simulations initialized with a helical sub-optimal perturbation of wavelength $1.5D$, as manifested in the experiments. In these simulations, a small amount of random noise was added to the helical perturbation to enable secondary (nonlinear) instabilities and turbulence breakdown \cite{SchmidHenningson_springer2001}. Indeed, after the initial development and amplification of the helical wave, breakdown to turbulence occurred. The peak in turbulent kinetic energy was reached at $t/T\approx0.75$, as shown in Fig.~\ref{fig:time_friction_energy}b, in close agreement with experiments. The strong fluctuations and abrupt changes in shear stress that occur during this period are shown in Fig.~\ref{fig:time_friction_energy}c. Again like in experiments, the fluctuations decayed during the acceleration phase and the flow returned to laminar. The helical vortex pattern in the radial-azimuthal plane and the waviness in radial-streamwise cross section resemble those in experiments, as shown in Fig.~\ref{fig:PIV_DNS_contour} (see also Movie $\mathrm{S}2\; \&\; \mathrm{S}3$ in \emph{SI}). It can hence be concluded that the large transient amplification of disturbances during flow deceleration provides a generic mechanism for the generation of helical vortices and a subsequent breakdown into turbulence.

In a recent investigation, Pier and Schmid \citep{Pier17} studied in detail how pulsation modifies the classic linear instability of channel flow (two-dimensional Tollmien--Schlichting waves). In agreement with von Kerzcek \cite{kerczek_instability_1982} they found that pulsation leads to a modulation of the growth rate of Tollmien--Schlichting waves. More specifically, they noted strong modal transient growth during deceleration and decay during acceleration.  While this phase relationship is in very good agreement with the one observed here, pipe flow is linearly stable and hence does not support Tollmien--Schlichting waves. On the other hand, the linear instability of pulsatile pipe flow identified by Thomas \textit{et al}.\ \cite{Thomas11} occurs only when the oscillatory component is predominant, i.e.\ for parameters far from cardiovascular conditions. It stems from the thin Stokes layer near the pipe wall and it occurs at much larger pulsation amplitudes (and Reynolds numbers), and is two-dimensional (axisymmetric, $m=0$). Our non-modal transient growth analysis shows that the energy of all axisymmetric perturbations decays nearly monotonically. The helical instability revealed here occurs at moderate amplitudes and is rooted on the strong non-modal transient growth of helical (three-dimensional) perturbations and is thus distinct from those reported previously in the literature.

\subsection*{Lumen Constriction}
The cross sections of blood vessels frequently deviate from the idealized circular case, for example protrusions may arise during wound healing or stenosis formation. To test if the helical instability may also arise under such conditions, we replaced the curved pipe segment by a straight section that includes a local constriction in form of a spherical cap (up to $D/4$ in height and a base cap diameter of $2D$, see Fig.~\ref{fig:facility}c). For increasing Reynolds number at $(\Wo,\A)=(5,0.85)$, also in this case a helical vortex pattern was found during the flow deceleration (see Movie S$4$ in \emph{SI}). The helical wave was first observed $40D$ downstream of the protrusion. At its maximum amplitude the turbulent patch stretches approximately from $35D$ to $55D$ downstream from the spherical cap. 

In an earlier study Blackburn \emph{et al}.\ \cite{blackburn2008} investigated linear non-modal transient growth after a severe axisymmetric stenosis for steady and pulsatile flows. They found that non-axisymmetric disturbances with $m=1$ (however without helical structure, but consisting of a sinuous shear layer) amplify the most. We performed experiments with a slight axisymmetric constriction, but did not observe the helical instability. While the growth of perturbations shown by Blackburn \emph{et al}.\ \cite{blackburn2008} may be related to the mechanism reported here, their strong stenosis modifies the basic flow very substantially, which is in contrast to the small disturbances used here in experiments and simulations. 

To further test the robustness of the helical instability, we changed the waveform of the pulsatile driving. The idealized sinusoidal flow rate modulation was replaced by the waveform typically observed in the aorta \cite{Fraser08}. Experiments were carried out in the $20~\mathrm{mm}$ pipe and the flow parameters were $(Re_m, \Wo, \A)=(1100,10,0.8)$. Again the helical instability was observed during flow deceleration followed by relaminarization as the flow was accelerated.

\subsection*{Blood Flow Experiments}
While the experiments reported so far were carried out in water, we next used blood as the working fluid. Blood has non-Newtonian properties and is a dense suspension of blood cells (e.g., red blood cells take up approximately $40\%$ of the volume fraction). For the experiments we used a scaled down set--up with a pipe diameter of $4~\mathrm{mm}$ which otherwise followed the same working principle as the larger diameter pipe. To perturb the flow a curved section was introduced $185D$ from the pipe inlet. 
{Since blood is opaque and the flow structure can not be observed directly, we monitored the differential pressure downstream of the curved section (see Fig.~\ref{fig:time_friction_energy}d). Flows were deemed unsteady if deviations in pressure were larger than twice the background noise level of the sensor.} Like in the Newtonian flow also the pulsatile blood flow became unstable during flow deceleration, and a considerable drag increase was detected approximately $20D$ downstream of the curved pipe segment. During the acceleration the flow stabilized and returned to the laminar friction value. The instability threshold for blood flow is shown by the orange symbols in Fig.~\ref{fig:Re_A}. In this case the transition occurs at lower $Re_m$ than for water flows, however for blood flow a more strongly curved segment was used to perturb the flow and we would hence expect an earlier onset. For pulsation levels typical for the Aorta i.e. $A\approx0.94$, the Reynolds number threshold was as low as $800$ and hence much lower than the commonly assumed value of $2000$.
The measurements were repeated under comparable condition using a transparent Newtonian fluid (water), where again the deviation in pressure was used to determine the instability threshold and was found to coincide with the appearance of the helical wave (blue line in Fig.~\ref{fig:Re_A}).

\section*{Discussion and Conclusion}
In summary, we report a generic instability for pulsatile pipe flow that occurs for large pulsation amplitudes and precedes the normal turbulence transition. The helical vortex pattern characteristic for this instability sets in at unusually low Reynolds numbers. As shown weak curvature and modest pipe constrictions are sufficient to destabilize the laminar flow. It is interesting to note that the geometrical perturbations that appear to be most efficient in pulsatile flow, are inefficient in the context of steady pipe flow. Curvature in fact has a stabilizing effect under steady conditions \citep{Kuehnen15} and can even lead to relaminarization \citep{Sreenivasan83} at not too large $Re$. Constrictions on the other hand need to be very severe \citep{Durst85} in order to trigger puffs in steady flow. Our study hence shows that pulsatile flows are susceptible to qualitatively different and more subtle perturbations than steady pipe flows. Another characteristic of the identified mechanism is that the instability only occurs during part of the pulsation cycle, i.e. the deceleration, whereas acceleration relaminarizes the flow. This particular feature is shared with linear modal and non-modal mechanisms uncovered recently in pulsatile channel flow \citep{Pier17,Tsigklifis17}. Above findings hence suggest that pulsatile flows of sufficient amplitude, such as cardiovascular flows in large blood vessels, despite being linearly stable can periodically break down into bursts of turbulence. The responsible transition mechanism requires perturbations of finite amplitude as caused by geometrical deviations from the straight pipe case (e.g. bends or constrictions). In particular the resulting large shear stress changes in space and time (see Fig.~\ref{fig:time_friction_energy}c and Fig.~\ref{fig:wall_shear_stress_contour}) encountered during flow deceleration offer a possible cause for endothelial activation.

%\newpage

%\subsubsection*{Appendices}
%
%PNAS prefers that authors submit individual source files to ensure readability. If this is not possible, supply a single PDF file that contains all of the SI associated with the paper. This file type will be published in raw format and will not be edited or composed.

\matmethods{
%Please describe your materials and methods here. This can be more than one paragraph, and may contain subsections and equations as required. Authors should include a statement in the methods section describing how readers will be able to access the data in the paper. 
\subsection*{Experimental Methods}\label{sec:exp_method}
{Experiments were carried out in straight, rigid pipes of circular cross section: (1) a $12~\textrm{m}$-long acrylic pipe (inner diameter $D=7.18 \pm 0.02 ~\textrm{mm}$) results in a measurement length of $1300D$ and this pipe was used for flow visualization and for measurement of puff survival probabilities; (2) a glass pipe (in diameter $D=20 \pm 0.01~\mathrm{mm}$) was used for PIV measurement (see below); (3) another glass pipe (in diameter $D=4 \pm 0.01~\mathrm{mm}$) was used for the blood flow experiments. In each case the pipe segments were positioned and carefully aligned on a long aluminum profile. The pipe is connected through a trumpet shaped convergence section to a reservoir (see the nozzle in Fig.~\ref{fig:facility}a). The rear end of the pipe is connected to a piston system. The volume of the piston can provide approximately $3000$ to $20000$ advective time units for observation of approximately $15$ to $450$ pulsation cycles for the Reynolds number investigated.
The plunger of the piston is driven by a motor through a gearbox. The speed of the motor is precisely controlled by a PC with a National Instruments card. The piston bore and the plunger speed set the cross--section averaged flow speed in the pipe $U(t) = U_m + U_o\cdot \textrm{sin}(2\pi f \cdot t)$. For the entire parameter regime under investigation the pipe flow is laminar unless perturbations are employed.} 
{The temperature of the fluid was measured before the experiments in order to correct viscosity changes and hence to accurately determine the Reynolds number. For the blood flow a milliliter of fluid was stabilized with $40~\mathrm{units}$ of an anticoagulant agent (Sigma-Aldrich). The kinematic viscosity of the blood (at laboratory room temperature $20^{\circ}$C) was measured to be $\nu=8\pm2~\textrm{mm}^2/\textrm{s}$.}

{The perturbation method applied to generate turbulent puffs was as follows: A small amount of fluid, corresponding to approximately $2\%$ of the pipe flow rate, was injected through a $1~\mathrm{mm}$ hole in the pipe wall. The perturbation point was located $150D$ downstream from the pipe inlet to allow for a sufficient pipe entry length. The duration of the injection was adjusted through an electronically controlled valve to cover the same phase in all experimental runs. A light sheet was used to illuminate the mid-plane (radial-streamwise) of the pipe. The fluid was seeded with fishsilver flakes for flow visualization. A digital camera (MatrixVision BlueFox 121G) was placed $1300D$ downstream from the injection point to record whether puffs decayed or survived. In each individual run, only one puff was generated in the pipe. $150$ runs were carried out for each selected Reynolds number (keeping the pulsation amplitude and frequency fixed) to give a reasonably well--converged survival probability of the puffs. }

{To trigger the helical instability, two perturbation methods as sketched in Fig.~\ref{fig:facility}(b, c) were used and they were produced using a three-dimensional printer. The ends of the perturbation sections were further finished in a milling machine to ensure a smooth connecting with the adjacent pipe segment. The curved pipe segment (see Fig.~\ref{fig:facility}b) is of cosinusoidal shape, and has the same inner diameter as the pipe. The constriction perturbation (see Fig.~\ref{fig:facility}c) is straight and has a protrusion in form of a spherical cap which is extended in the streamwise direction by $2D$. Its height ranges from $0$ to $D/4$. For both perturbations, the perturbation level is given by the offset $L_o$ divided by the corresponding length $L_p$.

For this set of experiments, a V$10$ Phantom high-speed camera (in resolution of $2400 \times 1800$ pixels$^2$) was used to visualize the helical instability. It was placed approximately $20D$ downstream of the perturbation section to record the flow and it was run at sampling rates up to $30$ frames per second. At the same position, the pressure drop was measured across a streamwise distance of $40D$ using a high-sensitivity differential pressure sensor (HSC series, Honeywell) with a sampling rate of $50~\mathrm{Hz}$.} 

{The velocity fields recorded during the occurrence of the helical instability were obtained by particle image velocimetry (PIV) measurements. The data were recorded in the $20~\textrm{mm}$ glass pipe. Two-dimensional planar PIV measurements were carried out in the mid cross section (radial-streamwise) of the pipe. To obtain all three velocity components, stereo-PIV measurements were carried out in the cross section perpendicular to the pipe axis (radial-radial). The measurements were performed approximately $20D$ downstream of the perturbation section. For these measurements the fluid (i.e. water) was seeded homogeneously with hollow-glass spheres which have a diameter of approximately $10$~$\mathrm{\mu m}$. The pipe cross section was illuminated using a continuous wave laser (center wavelength of $532~\mathrm{nm}$, FC $532$N-$5$W). A series of lenses was used to create a light sheet with a thickness of approximately $1~\mathrm{mm}$. A prism was used to minimize the imaging distortions that originated from the curvature of the pipe wall. The images were captured using Phantom V$10$ cameras. Commercial software DaVis (LaVision) was used to compute the velocity vectors through a multi-step algorithm. A $32\times32$ pixel window size with $50\%$ overlap was set for the final step for both sets of the PIV measurements.}

\subsection*{Numerical Methods}\label{sec:num_method}

We numerically computed the motion of an incompressible Newtonian fluid driven through a circular straight pipe at a pulsatile flow rate. In the axial direction, periodic boundary conditions were considered. The Navier--Stokes equations were rendered dimensionless by scaling lengths and velocities with the pipe diameter $D$ and the mean velocity $U_m$, respectively. Consequently,  time was rendered dimensionless by scaling with the advective time unit $D/U_m$. The instantaneous Reynolds number is $Re(t)=Re_m\cdot[1+\A\cdot \mathrm{sin}(2\pi t/T)]$,  where the dimensionless pulsation period is $T=\pi Re_m/(2\Wo^2)$.

For the linear analysis, we employed the adjoint-based method of Barkley \textit{et al.} \cite{Barkley08} to calculate the optimal growth for our system. Note however that in our problem the base flow is time-dependent, $\bm U_b(t)$, and is analytically given in ref.~\citenum{Womersley55}. The \emph{linearized} Navier--Stokes equations read 
\begin{equation}\label{equ:LNS}
\frac{\partial \bm u'}{\partial t}+{\bm u'}\cdot\bm{\nabla}{{\bm U}_b} 
+{\bm U}_b\cdot\bm{\nabla}{\bm u'} =-{\bm{\nabla}p'}+\frac{1}{Re_m}{\bm\nabla}^2{\bm u'}, \;
\bm{\nabla}\cdot{\bm u'}=0
\end{equation}
and the adjoint system reads
\begin{equation}\label{equ:LNS_adjoint}
\frac{\partial \bm u^*}{\partial t}-{\bm u^*}\cdot(\bm{\nabla}{{\bm U}_b})^\mathrm{Tr} 
+{\bm U}_b\cdot\bm{\nabla}{\bm u^*} ={\bm{\nabla}p^*}-\frac{1}{Re_m}{\bm\nabla^2}{\bm u^*}, \;
\bm{\nabla}\cdot{\bm u^*}=0.
\end{equation}
Here $\bm u'$ is a small velocity fluctuation with respect to the base flow ${\bm U_b}(t)$ and $p'$ is the pressure fluctuation. Starred quantities are the adjoints of the primed variables and $\mathrm{Tr}$ denotes matrix transpose. In the radial direction, no-slip boundary conditions were imposed for both $\bm u'$ and $\bm u^*$.

In pulsatile flow, the laminar base flow is time dependent and hence the transient growth depends on the time $t_0$ at which the disturbance is applied. For a perturbation applied at $t=t_0$, the optimal growth of the kinetic energy $E$ at time $\tau$($>t_0$) is defined as 
\begin{equation}\label{equ:TG_definition}
G(t_0,\tau)=\underset{{\left\|\bm u'(t_0)\right\|_2\neq0}}{\text{max}}\frac{E(\tau)}{E(t_0)},
\end{equation} 
where $\bm u'(t_0)$ is the initial perturbation to the base flow at $t=t_0$, i.e.\ $\bm U_b(t_0)$. $G(t_0,\tau)$ can be calculated as the largest eigenvalue of the operator $A^*(\tau)A(\tau)$, where $A(\tau)$ and $A(\tau)^*$ are the action operators that map $\bm u'(t_0)$ to $\bm u'(\tau)$ according to Eq.~\ref{equ:LNS} and $\bm u^*(t_0)$ to $\bm u^*(\tau)$ according to Eq.~\ref{equ:LNS_adjoint}, respectively. Operationally, this method integrates Eq.~\ref{equ:LNS} forward from $t=t_0$ to $t=\tau$ and Eq.~\ref{equ:LNS_adjoint} backward from $t=\tau$ to $t=t_0$. Subsequently, the Krylov subspace method is used to approximate the largest eigenvalue 
of $A^*(\tau)A(\tau)$. This procedure is iterated until the eigenvalue is sufficiently converged. 

We solved the linearized equations using a Chebyshev-Fourier-Fourier spectral method, in which velocity and pressure are represented as
\begin{equation}\label{equ:Fourier}
B(r,\theta,z,t)_{(k, m)}={\hat B}_{(k,m)}(r,t)e^{i(kz+m\theta)} + \mathrm{cc.},
\end{equation}
where $k$ (real number) and $m$ (integer) are the axial and azimuthal wavenumbers, respectively, ${\hat B}_{(k,m)}$ is the Fourier coefficient of the mode $(k,m)$ and $\mathrm{cc.}$ represents the complex conjugate.
The integration in time was performed using a second-order-accurate Adams-Bashforth/backward differentiation scheme and the incompressibility condition is imposed using a projection method \citep{Hugues98}. We used a time-step size $\Delta t=0.025$ and $96$ Chebyshev-Guass-Labatto grid points in the radial direction. The  analysis was performed using Matlab scripts based on those of ref.~\citenum{Trefethen00}.

A multi-parameter optimization process was carried out using the adjoint analysis. We computed the optimal growth at time $t$, $G(t)$, by optimizing over disturbance shape ($k\in[0, 2\pi]$ and $m=0,1,2,3$)
and time at which the disturbance was applied, $t_0$. In pulsatile pipe flow, the classic streamwise invariant optimal perturbation of steady pipe flow (with $(k,m)=(0,1)$) yields maximum $G(t)\approx800$ at $t/T\approx0.55$, see the red dotted line in Fig.~\ref{fig:time_friction_energy}c). Helical perturbations $(k\neq0,m=1)$ start to dominate from $t/T\approx0.62$, with the mode $(k,m)=(2\pi/3,1)$ yielding maximum $G(t)\approx 4\times10^4$ during the deceleration phase at $t/T\approx0.88$. 

{In addition, we carried out direct numerical simulations of the nonlinear Navier--Stokes equations in cylindrical coordinates $(r,\theta,z)$ using the `openpipeflow' code \citep{openpipeflow}. The code uses primitive variables and a pressure Poisson equation formulation with an influence-matrix technique. In the radial direction, spatial finite-difference discretization is employed with nine-point stencils, and points are densely clustered close to the pipe wall for capturing small flow structures. No-slip boundary conditions are applied at the pipe wall. Spectral methods are employed along the pipe axis ($z$) and azimuthal ($\theta$) direction to present periodicity, and the variables are expanded in Fourier modes 
\begin{equation}
V(r,\theta,z)=\sum_{k=-K}^{K}\sum_{m=-M}^{M}\hat{V}_{(k,m)}(r)\mathrm{e}^{i(\alpha kz+m\theta)}
\end{equation}
where $\hat{V}_{k,m}$ is the complex Fourier coefficient of the mode $(k,m)$ and $L_z=2\pi/\alpha$ is the pipe length.
The simulations were carried out at $(Re_m,\Wo,\A)=(2200,5.6,0.85)$ with $96$ radial points, $\pm 96$ and $\pm 196$  Fourier modes in the  azimuthal and axial directions for an approximately $12D$-long pipe. 
The Fourier modes (except for those corresponding to the optimal helical perturbation) were initialized with small values to mimic background noise in the experimental setup.} 

\subsection*{Data availability}\label{sec:data_availability}
The data can be found in Datasets S1--S8 in \emph{SI}.
}
\showmatmethods % Display the Materials and Methods section

\acknow{This work was supported by the Deutsche Forschungsgemeinschaft and the Austrian Science Fund in the framework of the research unit FOR 2688 ``Instabilities, Bifurcations and Migration in Pulsatile Flows'' Grant AV 120/6-1 and I4188-N30. D.X. gratefully acknowledges the support from Alexander von Humboldt Foundation (3.5-CHN/1154663STP). A.V. acknowledges support from the European Union's Horizon 2020 research and innovation programme under the Marie Sk{\l}odowska-Curie Grant 754411. B.S. acknowledges the support from the National Natural Science Foundation of China under Grant 91852105. We thank Davide Scarselli for his help with the PIV measurements.}

\showacknow % Display the acknowledgments section

% Bibliography
\bibliography{Xu2020arXiv}

\end{document}

\maketitle

\section*{Supporting Information (SI)}

\subsection*{Wall shear stress}

\begin{figure}
	\centering
	\includegraphics[width=0.6\linewidth]{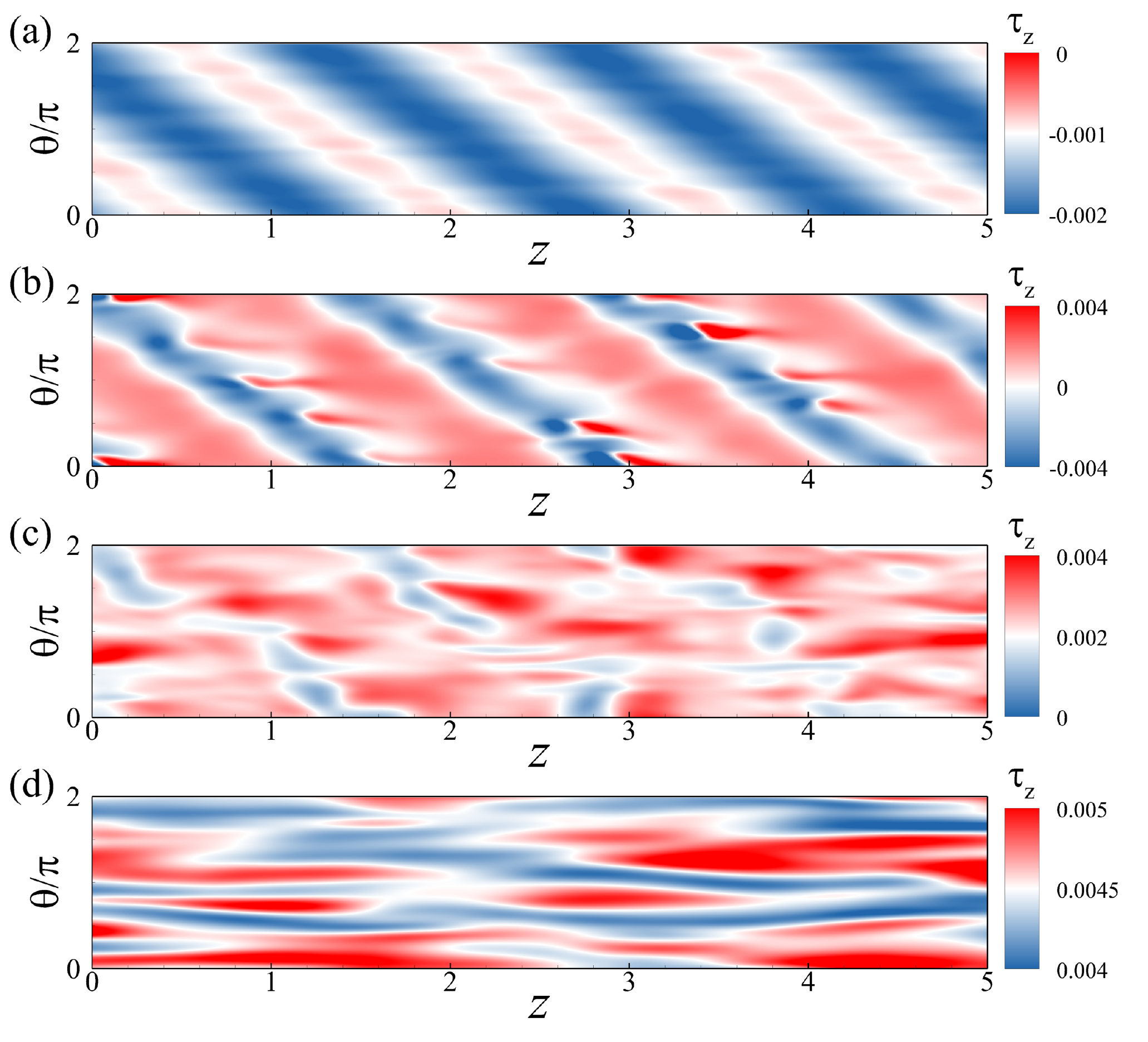}
	\caption{(Color online) The contour of wall shear stress $\tau_z$ from the numerical simulation at $(\Rey_s,\Wo,\A)=(2200,5.6,0.85)$ is shown for $t/T\approx 0.69$ (a), $0.77$ (b), $0.85$ (c) and $0.94$ (d). Note the contour range for panels differs for the highlight of flow patterns.}	
	\label{fig:wall_shear_stress_contour}
\end{figure}
Endothelial cells are shear sensitive \cite{nerem_role_1980, Glagov88, cunningham_role_2005} and high shear stress levels and steady flow conditions are considered to be atheroprotective \cite{Ku85, Hahn09}. Inversely disturbed, oscillating flow conditions and low shear stress levels cause inflammatory activation, endothelial dysfunction and may lead to the formation of atherosclerotic lesions \cite{cunningham_role_2005, gimbrone_endothelial_2016}. As shown below the helical instability identified in this study severely disrupts the regular shear stress level of laminar pulsatile flow. As shown in Fig.~\ref{fig:wall_shear_stress_contour} the growth of the helical perturbation during flow deceleration gives rise to regions of low shear stress and oscillations of the shear stress direction in space and time. The initially regular pattern of regions of alternating shear stress direction breaks down and becomes irregular as time proceeds and the flow turns turbulent (Fig.~\ref{fig:wall_shear_stress_contour} c to d).

\subsection*{Effect of nonlinearity on perturbation growth}
\begin{figure}
	\centering
	\includegraphics[width = 0.5\linewidth]{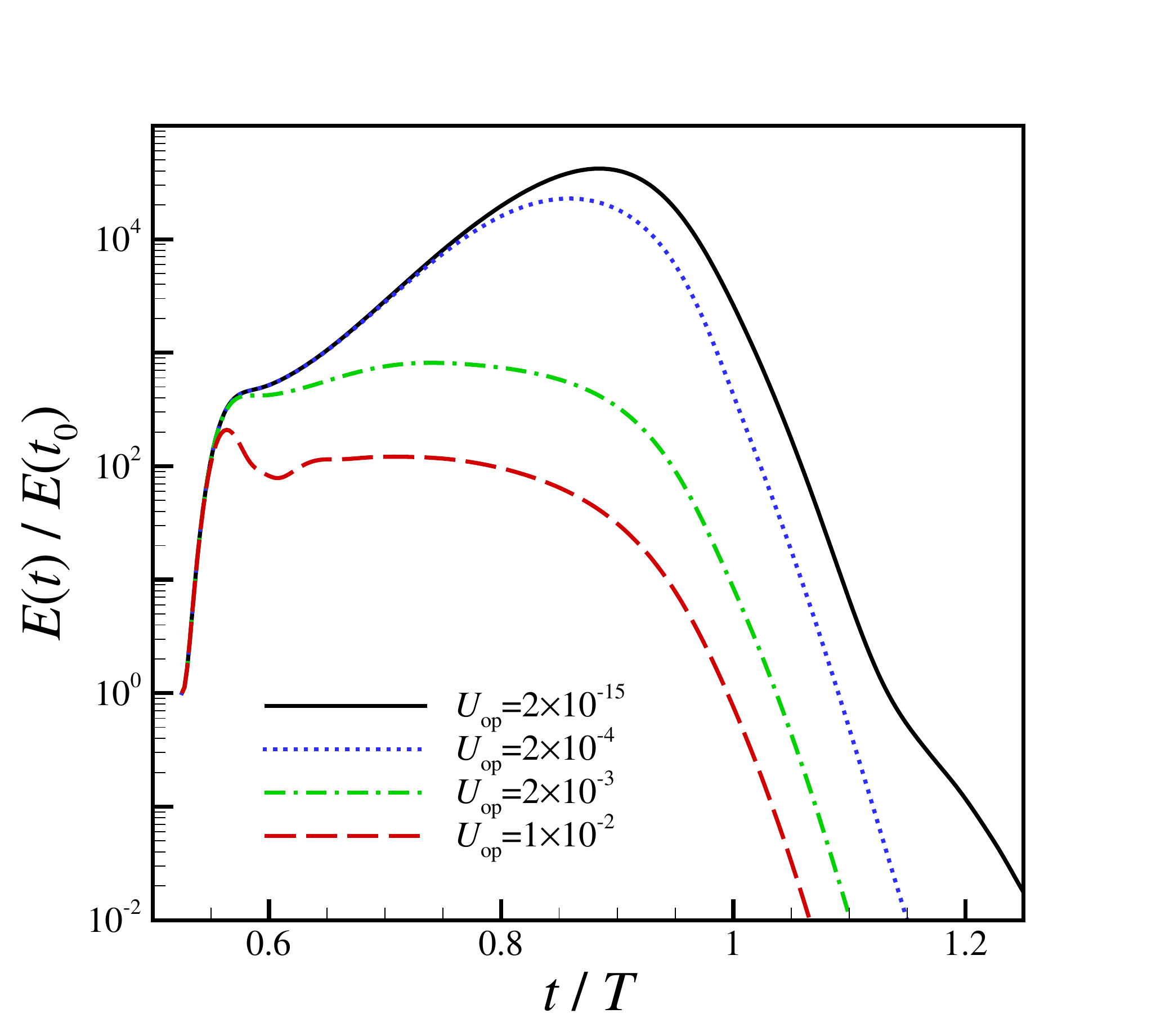}
	\caption{\label{fig:op_time_energy} {(Color online) The energy growth $E(t)/E(t_0)$ of the optimal perturbation $(k,m)=(2\pi/3,1)$ for four non-dimensional amplitudes in perturbation velocity $U_\mathrm{op}$ from $2\times10^{-15}$ to $1\times10^{-2}$. Here the energy $E(t_0)$ at the initial time $t_0$ is used to normalize the energy time series $E(t)$.}}
\end{figure}

Direct numerical simulations of the Navier--Stokes equations, including both linear and non-linear terms, were initialized at $(\Rey_s,\Wo,\A)=(2200,5.6,0.85)$ with an optimal perturbation i.e. at $(k,m)=(2\pi/3,1)$, in velocity $U_\text{op}=(E_\text{op}/E_{00}(t/T=0))^{1/2}$ starting from $2\times10^{-15}$. Reminded that $E_\text{op}$ is the energy of the perturbation and $E_\text{00}(t/T=0)$ is the energy of the laminar base flow at $t/T=0$ of the sinusoidal flow rate. Here no random noise was added. In Fig.~\ref{fig:op_time_energy}, the perturbation in the smallest amplitude (black solid line) gives purely linear energy growth up to the peaked $E(t)/E(t_0)\approx 4\times10^4$ which is the same as $G(t)$ given in the linear stability analysis. After the peak in $E(t)/E(t_0)$, the energy decays approximately exponentially. When the perturbation amplitude is increased, the initial state of the energy growth remains until $E(t)/E(t_0)\approx 2\times10^2$, in following the curves deviate from the linear trajectory. For $U_\text{op}=1\times10^{-2}$, the peaks of $E(t)/E(t_0)$ reduces to approximate $10^2$. This suggests that non-linear effects strongly quench the transient growth whilst shifting the energy peak to earlier times, which is consistent with previous studies \cite{Shi17}. Note however that nonlinear transient growth computations generally lead to more efficient disturbances than those obtained from linear analysis \cite{kerswell2014optimization}.

\subsection*{Turbulence detection in experiments}
\begin{figure}[htbp]
	\centering
	\includegraphics[width = 0.5\linewidth]{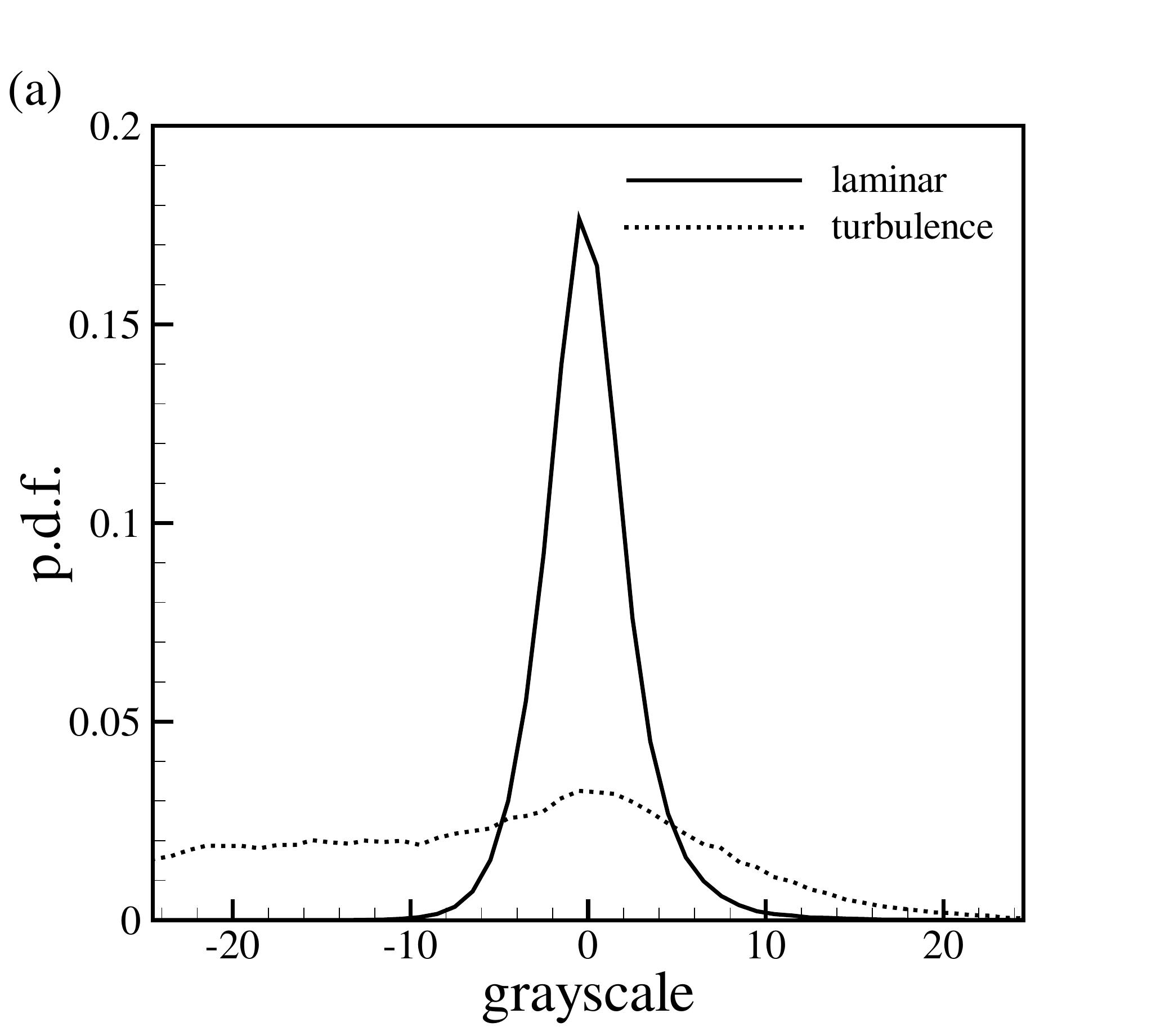}
	\includegraphics[width = 0.5\linewidth]{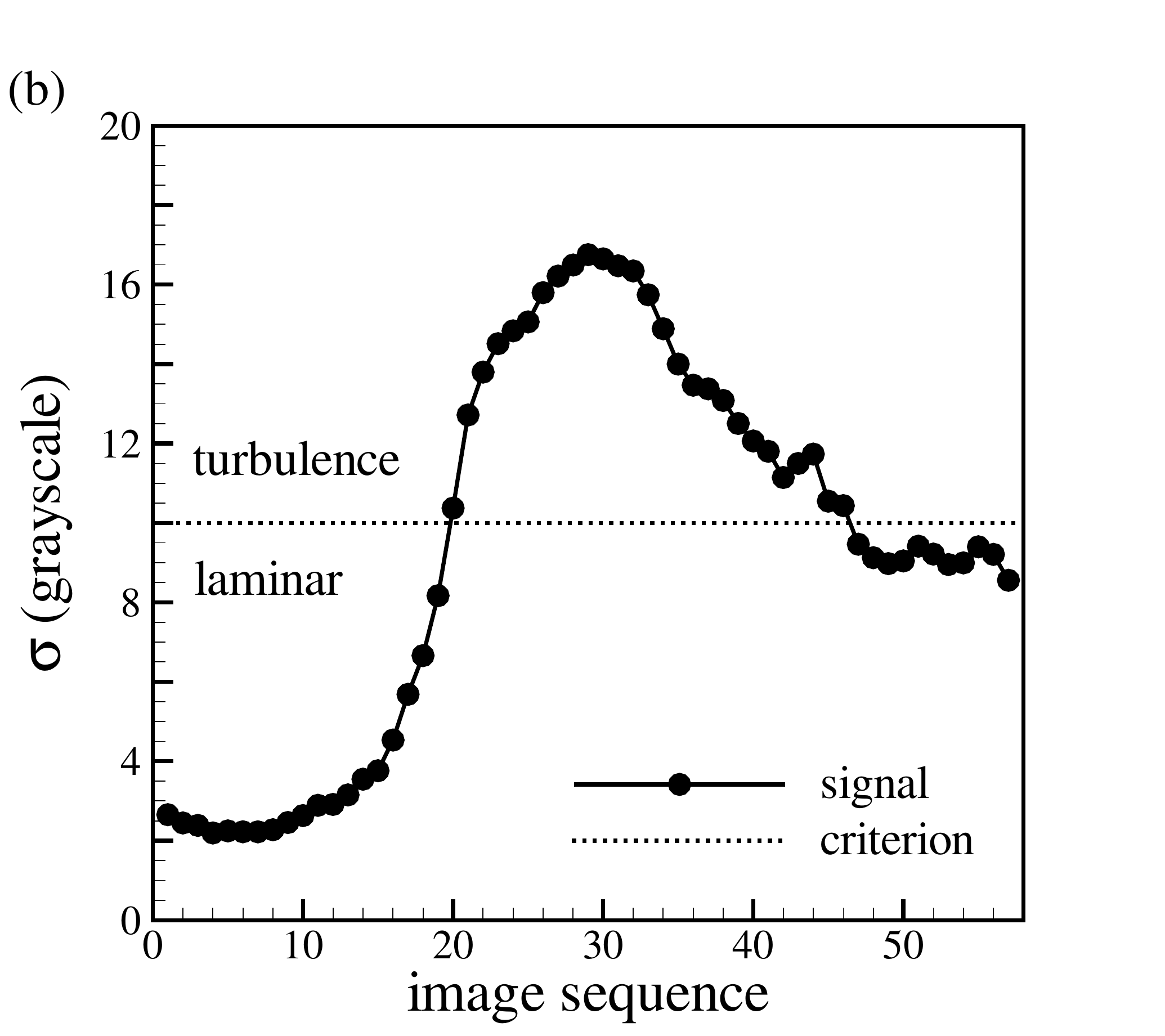}
	\caption{\label{fig:grayscale_pdf} {(a) The probability density function (p.d.f.) of the grayscale values of the measurement images after removal of the (laminar) reference signal. (b) The image sequence of the standard deviation ($\sigma$) of the image grayscale values with the reference signal removed.}}
\end{figure}

To detect turbulence in the experiments, the fluid was seeded with fishsilver flakes and the camera was used to record the flow structures. When the flow is laminar, the grayscale values of the images are uniform, whereas the grayscale values vary strongly in space when the flow is turbulent. Initially images of laminar flow were recorded to be averaged over the time series and the resulted image is used as the base reference. Once the images of the measurements were recorded, the base image was subtracted from each of those images. When the measured flow is laminar, the probability density function (p.d.f.) of the grayscale values is shown as a solid line in Fig.~\ref{fig:grayscale_pdf}a. When the flow is turbulent, the p.d.f. of the grayscale values covers a wider range with a small peak. Accordingly, the emergence of turbulence can be clearly identified from standard deviation ($\sigma$) of the grayscale values. Here a robust criterion $\sigma=10$ was used. In a recording (i.e. Movie S$1$ in \emph{SI}), as shown in Fig.~\ref{fig:grayscale_pdf}b, the flow is laminar indicated by small $\sigma$ then followed by an increase of $\sigma$. When $\sigma$ is above the threshold, the helical flow structures emerge and at the maximum of $\sigma$ the structures break down into turbulence. Subsequently the turbulence relaminarizes and is flushed downstream to be out of the field of the view.  

%%% Add this line AFTER all your figures and tables
\FloatBarrier

\movie{The helical instability in the experiment at ${(\Rey_m,\Wo,\A)=(2200,5.6,0.85)}$. The helical wave appears during flow deceleration, breaks down into disordered motion and eventually the flow relaminarizes during acceleration.}

\movie{Two dimensional PIV velocity vectors of the helical vortex pattern at the axial-radial cross section in the experiment at ${(\Rey_m,\Wo,\A)=(2200,5.6,0.85)}$. The vectors are color coded by the spanwise vorticity.}

\movie{Isosurfaces of streamwise vorticity in the direct numerical simulation showing the helical instability at ${(\Rey_m,\Wo,\A)=(2200,5.6,0.85)}$. Red and blue denote positive and negative values, respectively.}

\movie{Helical instability and subsequent break down into turbulence downstream of the lumen constriction at ${(\Rey_m,\Wo,\A)=(2200,5,0.85)}$.}

\dataset{datasetS1.txt}{The data for producing figure2.}

\dataset{datasetS2.txt}{The data for producing figure4a.}

\dataset{datasetS3.txt}{The data for producing figure4b.}

\dataset{datasetS4.txt}{The data for producing figure4c.}

\dataset{datasetS5.txt}{The data for producing figure4d.}

\dataset{datasetS6.txt}{The data for producing figure6.}

\dataset{datasetS7.txt}{The data for producing figureS2.}

\dataset{datasetS8.txt}{The data for producing figureS3.}

\bibliography{Xu2020arXiv}